\documentclass[aps,superscriptaddress,showpacs,twocolumn,amsmath]{revtex4}
\usepackage{caption,graphicx,epsfig,epstopdf}
\usepackage{hyperref}
\usepackage{color}
\begin{document}

\title{Structure of proton based on the classical string model}
\author{Z. G. Tan }\thanks{tanzg@ccsu.edu.cn}\author{Y. F. Mo}\author{L. P. Sun}
\affiliation{Department of Electronic and Communication Engineering,Changsha University,Changsha, 410003,P.R.China}

\begin{abstract}
In this paper, we propose several proton structure imagination scenarios based on classical string model. The radius and mass properties of protons in relativistic and non-relativistic cases are discussed, Contrary to asymptotic freedom, we find that the interaction between quarks in protons is inversely proportional to distance.
\end{abstract}

\maketitle

\section{Introduction}\label{sec1}
Exploring the structure of basic particles that make up the universe and their interactions is one of the main tasks of fundamental research in physics. With the increase of collision energy of the Large Hadron Collider (LHC), the understanding of the microscopic structure of matter has penetrated into nucleon.  A standard model of particle physics is established at the quark level. The model considers that all matter consists of a small number of basic fermions with half-integer spins, including three generations of quarks and three generations of leptons. The bosons whose spins are integers exist as the medium of interaction between fermions. Mesons, for example, consist of a pair of positive and negative quarks, and baryons usually consist of three constituent quarks. Quarks are strongly interacting with each other through gluons with a very short transmission range. Mesons and baryons are collectively called hadrons. The study of the internal structure of hadrons, i.e. the configuration distribution of these quarks in hadrons, has become a hot topic in  high energy physics\cite{c1,c2,c3}.Although protons and neutrons are considered to be the basic components of visible matter, their internal structures are not simple. They are actually complex bound states of quarks and gluons\cite{c4,c5}. Strong interactions described by quantum chromodynamics (QCD) are combined. The momentum distribution of quarks and gluons in hadrons can be fitted by analyzing the momentum distribution of the final particles in the high energy collision experiment data, and some experimental expectations are given\cite{c6}. The hadron bag model\cite{c7} equates the effect of gluon to a bag that restricts the escape of quarks, which are trapped in the bag for thermal motion. Through high-energy collision, the hadron bag was broken, and the quarks were released briefly. Subsequently , the quarks quickly recombined to form a new Hadron bag. The string model\cite{c8} regards hadrons as strings with quarks at both ends. The strings are torn apart by the enormous energy of collision experiments, so that new quarks and gluons are "torn" out of the vacuum, forming new bound states (hadrons) around quarks.

For a long time, neither bag model nor string model can accurately describe the proton internal dynamics\cite{c9}. On the one hand, QCD theory itself is not easy to generate any real-time images using Euclidean space representation, on the other hand, because in such a small range, any concept of orbit and momentum structure do not have any practical significance. However, just like the structure of hydrogen atoms, although quantum mechanics shows that electrons do not move around the nucleus in a fixed orbit, a simple nuclear structure model can still give many intuitive images and some basic particle properties\cite{c10}. Therefore, people still have strong expectations for the simple geometric structure of protons. This article is an exploration under this expectation. The organization of this paper is as following, In the next section, we propose several possible proton structure imagination scenarios according to classical mechanics and string theory. In Sec.~\ref{sec3}, we discuss the proton eigenvalues in several structures under relativity. Finally, there is a brief conclusion.

\section{Proton Structure Imagination without relativistic effect}\label{sec2}
Before conceiving the proton internal structure diagram, we propose the following basic assumptions
   \begin{enumerate}
   \item The interaction between quarks in hadrons is expressed by strings which strength  parameter a, called string tension, can be used as "rest mass density" of the string.
   \item Quarks must make steady circular motion inside hadrons, and their centripetal force comes from the tension of strings.
   \item Since quarks do not distinguish between tastes, quarks in protons should have the same status, that is, they have taste symmetry.
   \end{enumerate}

\begin{figure}[htb]
\center{\includegraphics[width=0.45\textwidth]{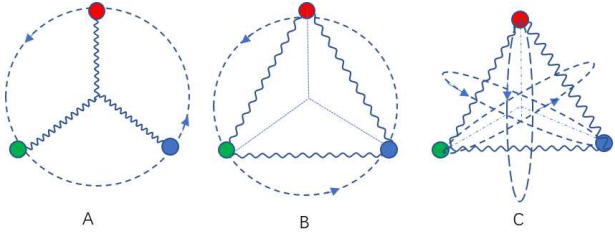}}
\caption{Three possible pictures of proton structure do not distinguish quark masses, and strings represent gluons. }\label{Fig1}
\end{figure}

For protons, if you don't distinguish between $u $ quark and $d $ quark, the possible internal picture is shown in FIG.\ref{Fig1}. In the first two pictures, three quarks that make up the proton are pulled by strings to rotate in a plane determined by them. In the latter picture, there are three rotating planes, each quark moves around the center of mass.
In the absence of relativity, the kinetic equation of quark motion is
\begin{eqnarray}
  m_q\omega^2R &=& a,\quad \mbox{(for A)}\\
  m_q\omega^2R &=& \sqrt{3}a. \quad \mbox{ (for B\&C) }
\end{eqnarray}
In the formula above, $m_q$ is quark mass, $m_q = 5MeV$, and $R$ is radius of proton. The string tension, for illustrative estimation, we will use experimental values
\[a=0.176GeV^2\approx 1GeV/fm,\]
as determine, for instance,from the spectra of quarkonia\cite{c11,c12}. In this way, the proton mass can be obtained.
\begin{eqnarray}
  M_p &=& 3aR+3m_q+3E_{kq} \nonumber \\
   &=&3aR+3m_q+\frac{3}{2}m_q(\omega R)^2 \nonumber \\
   &=& \frac{9}{2}aR+3m_q,\quad \mbox{(for A)}  \\
   &=& \frac{9\sqrt{3}}{2}aR+3m_q,\quad\mbox{(for B\&C)}
\end{eqnarray}

\begin{figure}[htb]
\center{\includegraphics[width=0.45\textwidth]{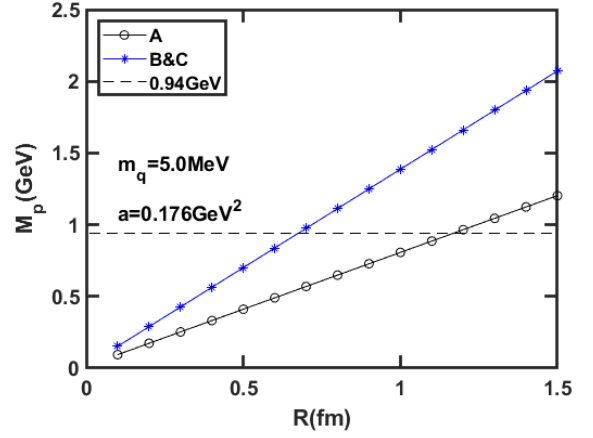}}
\caption{In the non-relativistic case, the relation curve between proton mass and radius. }\label{Fig2}
\end{figure}

FIG. \ref{Fig2} describes the relationship between proton mass and radius in non-relativistic cases. In the case of a recognized proton mass of $M_p\approx 0.94 GeV$. the proton radii corresponding to structure A and B are respectively determined $1.17fm, 0.67fm$.

\section{Proton structure with relativistic effect}\label{sec3}
Considering the relativistic effect, for structure A, the string length is exactly the radius $R$. For circular motion, the radius is perpendicular to the direction of motion, so it will not shrink. But the circumference of the circle will shorten.
\begin{equation}L=2\pi R\sqrt{1-v^2}\quad \mbox{unit} c=1.\label{e1}\end{equation}
Thus, the rotation pattern becomes

\begin{figure}[htb]
\center{\includegraphics[width=0.25\textwidth]{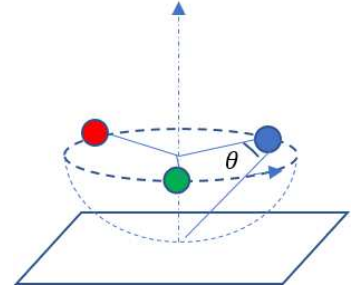}}
\caption{Circumferential Motion Shape Considering Relative Effect }\label{Fig3}
\end{figure}
\begin{figure}[htb]
\center{\includegraphics[width=0.45\textwidth]{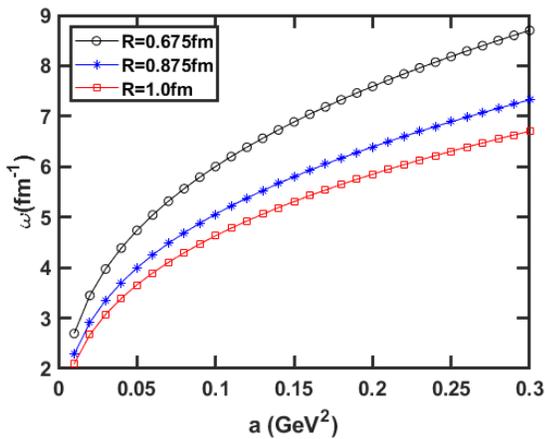}}
\caption{The relation between the angular velocity of the quark rotating around the center of mass and the tension of the string  }\label{Figomega}
\end{figure}

In Eq.(\ref{e1}), $v$ is not equal to $\omega R$, but should be $ \omega r$. The relationship between $r$ and $R$ can be derived from the following equation
\begin{eqnarray}
2\pi r&=&2\pi R\sqrt{1-\omega^2 r^2}\nonumber\\
r^2&=&R^2(1-\omega^2 r^2)\nonumber\\
\frac{r}{R}&=&\frac{1}{\sqrt{1+\omega^2 R^2}} \label{e2}
\end{eqnarray}
Let the quark mass be $m_q$ and the string tension be $a$, Then the following dynamic relations are established.
\begin{eqnarray}
\frac{m_q\omega^2r}{\sqrt{1-\omega^2 r^2}}&=&a\cos\theta=a\frac{r}{R}\nonumber\\
&=&a\frac{1}{\sqrt{1+\omega^2 R^2}}=a\sqrt{1-\omega^2 r^2}
\end{eqnarray}
Thus, given $a, m_q, R$ the $\omega$ can be solved by the following equation
\begin{equation}
m^2R^4\omega^6+m^2R^2\omega^4-a^2=0
\end{equation}

set $\omega^2=W$, We get \begin{equation}\label{ew3}
 m^2R^4W^3+m^2R^2W^2-a^2=0\end{equation}
For such a cubic equation of one variable with respect to W, because angular velocity must be real, the solution can be obtained by Cardans formula
\begin{eqnarray}
W&=&\sqrt[3]{\frac{a^2}{2m^2R^4}-\frac{1}{27R^6}+\frac{a}{mR^4}\sqrt{\Delta'}} \nonumber\\
&&+\sqrt[3]{\frac{a^2}{2m^2R^4}-\frac{1}{27R^6}-\frac{a}{mR^4}\sqrt{\Delta'}}-\frac{1}{3R^2}
\end{eqnarray}
where
\[\Delta'=\frac{a^2}{4m^2}-\frac{1}{27R^2}\]
Thus $\omega= \sqrt {W}$ is available. FIG.\ref{Figomega} shows the relationship between the angular velocity of quark rotating around the center of mass of proton and the tension of the string at several different radii. It can be seen from the Figure that the angular velocity increases with the string tension. For the same string tension, the smaller the radius, the larger the angular velocity.
With relativistic effect, the energy of a single string is
\begin{eqnarray}
  E_0 &=& m_s= \int_0^R\frac{adx}{\sqrt{1-\omega^2r^2}}=\int_0^Ra\sqrt{1+\omega^2x^2}dx \nonumber\\
   &=&\frac{a}{\omega}\int_0^{\omega R}\sqrt{1+(\omega x)^2}d(\omega x) \nonumber \\
   &=& \frac{a}{\omega}\left(\frac{\omega R}{2}\sqrt{1+\omega^2R^2}+\frac{1}{2}\ln(\omega R+\sqrt{1+\omega^2R^2})\right)\nonumber\\
   &&
\end{eqnarray}
The total mass of protons is the sum of string energy and quark mass
\begin{eqnarray}
  M &=& 3E_0+3m_q\sqrt{1+\omega^2R^2} \nonumber \\
   &=& 3\frac{a}{\omega}\left(\frac{\omega R}{2}\sqrt{1+\omega^2R^2}+\frac{1}{2}\ln(\omega R+\sqrt{1+\omega^2R^2})\right)\nonumber\\
   &&+3m_q\sqrt{1+\omega^2R^2}\label{eqmass}
\end{eqnarray}

\begin{figure}[htb]
\center{\includegraphics[width=0.45\textwidth]{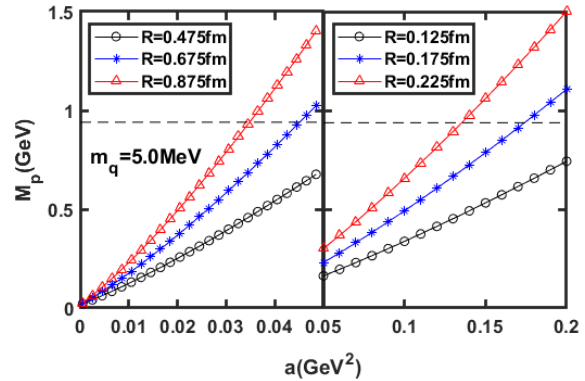}}
\caption{The relation between the mass of proton and the tension of the string  }\label{figmass}
\end{figure}

\begin{figure}[htb]
\center{\includegraphics[width=0.45\textwidth]{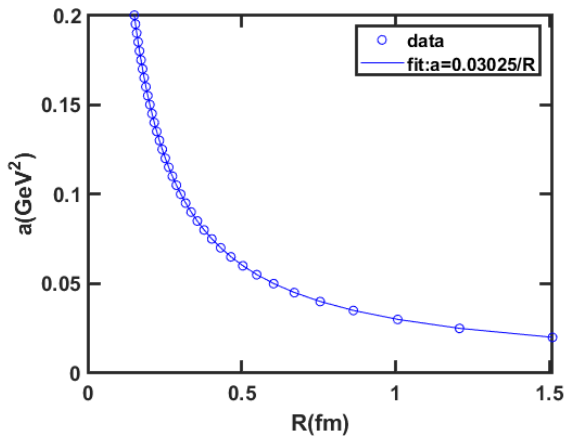}}
\caption{The the relationship between string tension and radius of proton }\label{figar}
\end{figure}

Eq.(\ref{eqmass})gives an explicit functional relationship between proton mass and $m_q$, string tension $a$ and radius $R$. As can be seen from FIG.\ref{figmass}, when the mass of quark and proton are fixed as $5MeV$ and $0.94GeV$, the radius of proton obviously depends on the tension of the string.
\[M_p=M_p(m_q,a,R)\]
As shown in FIG.\ref{Figomega}, if the string tension remains unchanged, the rotation angular velocity will increase with the decrease of radius. Because of the relativistic effect, the single string energy and the quark kinetic energy will also increase with the increase of rotational speed, which leads to the increase of proton mass. The proton mass measured by experiments is basically a constant. To meet this requirement, the string tension must be reduced.  To highlight this point, we describe the relationship between string tension and radius as shown in FIG.\ref{figar}. Where the small circle in the figure represents the data we have calculated by Eq.(\ref{eqmass}), and the real line is the function relationship fitted.
\begin{equation}\label{eqar}
  a=\frac{0.03025}{R}
\end{equation}

Structures B and C in FIG.\ref{Fig1} can be equivalent to A with a constant factor difference between them. So no further discussion.
\section{Conclusion}
   In this paper, We proposed a few structures of proton based on some classical imagine. Because quarks and strings move so fast, the relativistic effect cannot be ignored. However, Unlike the asymptotic freedom predicted by QCD theory, our calculations show that the tension of the string is inversely proportional to the proton radius and therefore to the distance between quarks. Although the discussion in this article is rather rough, this conclusion is still worth pondering.Perhaps the energy fluctuation caused by the longitudinal vibration of the string should also be considered, if possible, this is the next work, including angular momentum calculation, spin, etc.
\section*{Acknowledgement}
We are grateful to the financial support (grant No. 18C0762) from Department of Education of Hunan Province, China. This work was supported also, in part, by the Theoretical Physics Special Project of National Natural Science Foundation of China (Grant No. 11747123) and the Natural Science Foundation of Hunan Province of China (Grant No. 2018JJ3560,and 2018JJ2455).

\end{document}